*UDC 535.71*

# Continual Model of Medium IV: Calculating the Analytical Gradients of the Solvation Energy over the Atomic Coordinates

O. Yu. Kupervasser[a,*] and N. E. Wanner[b]
[a]TRANSIST VIDEO Ltd., Skolkovo resident
[b]All-Russia Research Institute of Veterinary Sanitary, Hygiene, and Ecology, Russian Academy of Agricultural Sciences, Moscow
*e-mail: olegkup@yahoo.com

**Abstract—**In this work, we describe some methods of finding the analytical gradients (derivatives) of the solvation energy over the atomic coordinates. This procedure is performed for both the non-polar component of the solvation energy and its polar components obtained by the PCM, COSMO, and SGB methods. These gradients are found using the analytical gradients of the parameters of surface elements (coordinates, normals, and surface areas). Surface elements lie on both an optimally smooth solvent excluded surface (SES) and a solvent accessible surface (SAS) obtained from SES. These surfaces are found via primary and secondary rolling by the algorithm described in [1–9].

**Key words:** solvation energy, analytical gradients, numerical gradients, molecular surface, primary rolling, secondary rolling

## 1. Introduction

The free solvation energy $\Delta G_s$ in continual solvent models is represented as the sum of the three following components [10]:

$$\Delta G_s = \Delta G_{cav} + \Delta G_{np} + \Delta G_{pol}, \quad (1)$$

$$\Delta G_{nonpol} = \Delta G_{cav} + \Delta G_{np},$$

where $\Delta G_{pol}$ is the polar component associated with the polarization of a dielectric solution, $\Delta G_{cav}$ is the cavitation component associated with the formation of a cavity in a solvent, $\Delta G_{np}$ is the component determined by van der Waals interaction, and $\Delta G_{nonpol}$ is the non-polar component representing the sum of the two previous components.

Let us give a brief review of the works that deal with finding the analytical gradients (derivatives) of the solvation energy over the atomic coordinates. Similar gradients are necessary, for example, for the final optimization of the position and configuration of a ligand molecule in the model of docking.

The problem of finding the analytical gradients is divided into the two subproblems:

(1) Finding the analytical gradients of the parameters of molecular surface elements (coordinates, normals, surface areas). This problem is considered in [11] for the GEPOL algorithm of constructing a surface (by filling the empty space with fictitious balls). The algorithm of constructing a smooth surface via primary and secondary rolling is described in [1–9].

(2) Calculating the analytical gradients of the matrices incorporated into the equation of a model on the basis of the above found gradients of the parameters of surface elements (***CO****nductor-like* ***S****creening* ***MO****del* (COSMO) [12–13], *Polarized Continuum Model* (PCM) [14–15]). The analytical gradients of the solvation energy are then calculated using the gradients of these matrices.

It should be noted that, for the PCM method, the technique of finding the gradients [14–15] requires one of the matrices in the PCM equation to be inverted. If the number of surface elements is high, the order of this matrix is also high. The storage of such a matrix requires a great deal of computer memory and considerably increases the computational time of the problem. Moreover, the matrix inversion algorithms themselves [16–17] also require a lot of memory and time. However, for the method with enlarged surface elements [18], the inverse matrix is small, and its calculation and use are not problematic. The inverse matrix can iteratively be found by the method [19]. Using the methods [14–15], we can find the analytical gradients themselves.

In the given work, we describe the solution of the second subproblem that consists in finding the analytical gradients for the polar component of the solvation energy for such methods as PCM [20], COSMO [12], and SGB [21], and also its non-polar component [20]. To solve this subproblem, we use the analytical gradients of the parameters of elements on molecular surfaces obtained by primary and secondary rolling [1–9].

There exist the two types of surfaces over a molecule [1–9]. First, this is the solvent excluded surface (SES). The volume occupied by a solvent lies *outside* the volume enveloped by this surface. A substrate itself lies completely *inside* this volume. The algorithm [1–9] forms a smooth SES via primary and secondary rolling. Second, this is the solvent accessible surface (SAS) formed by the centers of solvent molecules tangent to a substrate molecule in primary rolling. The first type of surface is used to calculate the polar component of the solvation energy (really, many molecule's sites important for electrostatic interaction have a zero or very small surface areas [22]), and the second type of surface is used to determine the cavitation and van der Waals components. The calculations of the analytical gradients of the energy obtained by the PCM, COSMO, and SGB methods are implemented in the DISOLV software [1–9].

## 2. Physical Models

### 2.1. Polar Component of the Solvation Energy

To calculate the polar component of the solvation energy, we take SES elements.

#### 2.1.1. Polarized Continuum Model

##### 2.1.1.1. Principal Equations for the Energy in the Polarized Continuum Model

Let us now give the detailed description of the models used in our work. Let us begin with PCM [16], [19–20] and consider the above described problem of finding the polar component of the solvation energy. We replace the solution $\Omega_{ex}$ by a dielectric with the known dielectric permittivity $\varepsilon_{ex}$. A dissolved molecule is considered as the dielectric's cavity $\Omega_{in}$ filled either with a vacuum ($\varepsilon_{in}$=1) or a dielectric with a dielectric permittivity $\varepsilon_{in}\neq 1$. The point charges $Q_i$ located in the centers of atoms $\boldsymbol{R}_i$ at a certain distance form the boundary $\Xi$ are inside this cavity. These charges that roughly approximate the distribution of charge inside a molecule were found by the Merck molecular force field (MMFF94) methods [23–25]. There is a possible situation when we deal with several cavities corresponding to several molecules instead of a single cavity. In this case, the effect of a dielectric solution is completely governed by the surface charge $\sigma(\boldsymbol{r})$ at the boundary of a cavity. The precise linear equation relating this surface charge with the position and value of charges inside a cavity will be called the PCM equation. To find the polar component of the solvation energy, we first calculate the energy $E_1$ necessary to transfer a molecule from a dielectric with $\varepsilon_{in}$ into a vacuum (the surface charge formed in this process has the density $\sigma_{vac}(\boldsymbol{r})$) and then the energy $E_2$ necessary to transfer a molecule from a dielectric with $\varepsilon_{in}$ into a dielectric with $\varepsilon_{ex}$. The solvation energy of the transfer of one mole of a substrate from a vacuum into a solvent with $\varepsilon_{ex}$ is equal to the difference between these two energies $\Delta G_{pol} = E_2 - E_1$ and is described by the following equation:

$$\Delta G_{pol} = \frac{1}{2}\sum_i Q_i \int_{\Xi} \frac{\sigma(\boldsymbol{r}) - \sigma_{vac}(\boldsymbol{r})}{|\boldsymbol{R}_i - \boldsymbol{r}|} dS` . \tag{2}$$

For $\varepsilon_{in} = 1$, the surface charge in a vacuum $\sigma_{vac} = 0$ and, consequently,

$$\Delta G_{pol} = \frac{1}{2}\sum_i Q_i \int_{\Xi} \frac{\sigma(\boldsymbol{r})}{|\boldsymbol{R}_i - \boldsymbol{r}|} dS` . \tag{3}$$

Since $\varepsilon_{in}$ is an empirically fitted parameter, we selected it as $\varepsilon_{in} = 1$ to exclude the vacuum component of the energy from consideration and simplify our calculations.

The integral PCM equation for the surface charge has the following form:

$$\sigma(\boldsymbol{r}) = \frac{\varepsilon_{in} - \varepsilon_{ex}}{2\pi(\varepsilon_{in} + \varepsilon_{ex})} \left( \sum_i \frac{Q_i\left((\boldsymbol{r} - \boldsymbol{R}_i) \cdot \boldsymbol{n}\right)}{\varepsilon_{in}|\boldsymbol{r} - \boldsymbol{R}_i|^3} + \int_{\Xi} \frac{\sigma(\boldsymbol{r}`)\left((\boldsymbol{r} - \boldsymbol{r}`) \cdot \boldsymbol{n}\right)}{|\boldsymbol{r} - \boldsymbol{r}`|^3} dS` \right). \tag{4}$$

The last integral is singular at $\boldsymbol{r} = \boldsymbol{r}`$ and, by definition,

$$\int_{\Xi} \frac{\sigma(\boldsymbol{r}`)\left((\boldsymbol{r} - \boldsymbol{r}`) \cdot \boldsymbol{n}\right)}{|\boldsymbol{r} - \boldsymbol{r}`|^3} dS` = \lim_{\delta \to 0} \int_{\Xi / (|\boldsymbol{r} - \boldsymbol{r}`| < \delta)} \frac{\sigma(\boldsymbol{r}`)\left((\boldsymbol{r} - \boldsymbol{r}`) \cdot \boldsymbol{n}\right)}{|\boldsymbol{r} - \boldsymbol{r}`|^3} dS` . \tag{5}$$

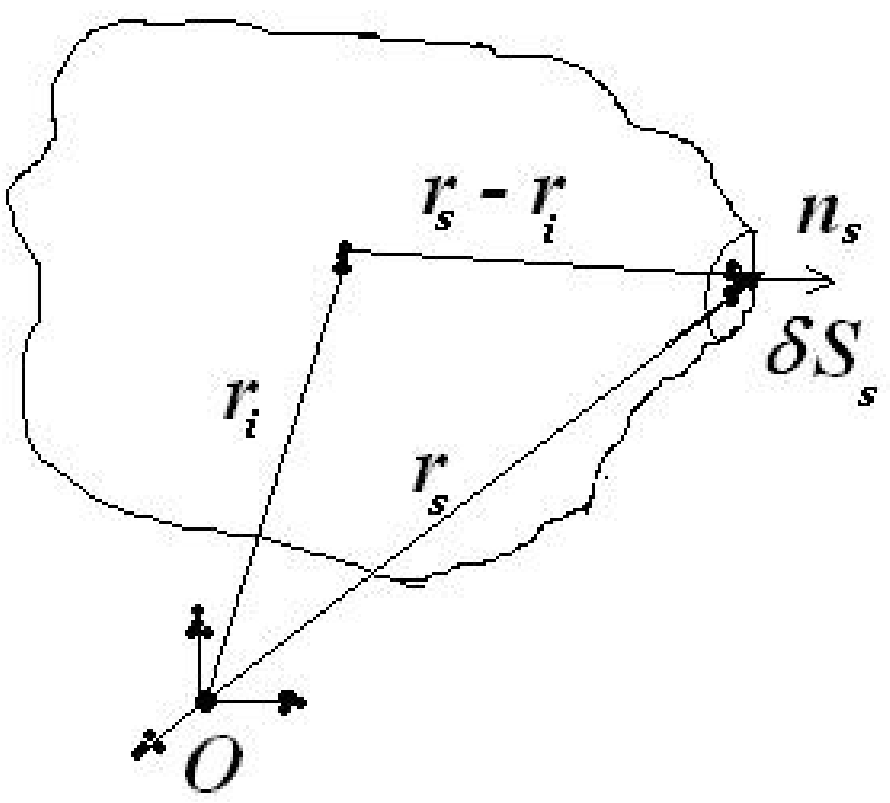

Fig. 1. Surface element relative to the origin of coordinates.

After the discretization of a surface (Fig. 1) (i.e., after its partitioning into small surface elements carrying surface charge), this linear equation may be written in the matrix form as

$$Aq = BQ, \tag{6}$$

where $A$ is the matrix depending on the parameters of surface elements, $B$ is the matrix depending on the geometric parameters of surface elements and the positions of charges inside a molecule, $q$ is the column of surface element charges, and $Q$ is the column of charges inside a molecule.

Let us determine the parameters of small surface elements and then determine how the elements of the above described matrices, such as the number of surface elements $M$, the number of charges inside a cavity $N$, the charge of the $j^{\text{th}}$ surface element $q_j$, the area of a surface element $S_j$, the radius vector of the center of a surface element $\boldsymbol{r}_j$, the normal at the center of a surface element $\boldsymbol{n}_j$, the charges inside a molecule $Q_i$, the radius vector of the $i^{\text{th}}$ charge inside a molecule $\boldsymbol{R}_i$, and $\varepsilon=\varepsilon_{out}/\varepsilon_{in}$, are expressed though them.

The elements $b_{ij}$ of the matrix $B$ are determined as

$$b_{ij} = \frac{\boldsymbol{n}_i(\boldsymbol{r}_i - \boldsymbol{R}_j)}{\|\boldsymbol{r}_i - \boldsymbol{R}_j\|^3} S_i \frac{1-\varepsilon}{4\pi(\varepsilon+1)} \frac{1}{\varepsilon_{in}}. \tag{7}$$

For the column of the matrix $B$, the following identity is true

$$\sum_{i=1}^{M} b_{ij} = \frac{1-\varepsilon}{1+\varepsilon}\frac{1}{\varepsilon_{in}}. \qquad (8)$$

The elements $a_{ij}$ of the matrix $A$ are expressed as

$$a_{ij} = \begin{cases} \dfrac{\boldsymbol{n}_i(\boldsymbol{r}_i - \boldsymbol{r}_j)}{\|\boldsymbol{r}_i - \boldsymbol{r}_j\|^3} S_i \dfrac{\varepsilon - 1}{4\pi(\varepsilon+1)} & i \neq j \\ a_{jj} = \dfrac{\varepsilon}{\varepsilon+1} - \sum_{i \neq j} a_{ij} & i = j \end{cases}. \qquad (9)$$

For the columns of the matrix $A$, the following identity is true

$$\sum_{i=1}^{M} a_{ij} = \frac{\varepsilon}{\varepsilon+1}. \qquad (10)$$

Expressions (8) and (10) automatically give the identity that relates the total surface charge with the total charge inside a cavity:

$$\sum_{j=1}^{M} q_j = -\left(\frac{1}{\varepsilon_{in}} - \frac{1}{\varepsilon_{ex}}\right)\sum_{i=1}^{N} Q_i. \qquad (11)$$

The use of Eq. (8) for the correction of numerical errors automatically leads to the very precise fulfillment of the above identity.

Each of the two energy components in Eq. (2) or the energy in Eq. (3) may be written in the matrix form as

$$\Delta G_{pol} = Q^T D q \qquad (12)$$

where the elements $d_{ij}$ of the matrix $D$ are determined as

$$d_{ij} = \frac{1}{2}\left(\frac{1}{\|\boldsymbol{R}_i - \boldsymbol{r}_j\|}\right). \qquad (13)$$

### 2.1.1.2. Analytical Derivatives for the Matrices in the Polarized Continuum Model

Let $R_{i,x^k}$ be the $x^k$ th component of the radius vector $R_i$, $x^k$ be the member of the set $\{x^0, x^1, x^2\}$, $i$ be the number of a shifted atom ($0 < i < N$), $\boldsymbol{e}^k$ be the basis unit vector in the direction of the coordinate $x^k$ and a member of the set $\{\boldsymbol{e}^0, \boldsymbol{e}^1, \boldsymbol{e}^2\}$, and $\delta_{ij}$ be the delta function.

The derivative of the elements of the matrix $A$ over the coordinate of one of the atoms $R_{m,x^k}$ is determined as

$$
\begin{cases}
\dfrac{\partial a_{ij}}{\partial R_{m,x^k}} = \dfrac{1}{4\pi}\dfrac{\varepsilon-1}{\varepsilon+1}\left(
\begin{array}{l}
\dfrac{\boldsymbol{n}_i\cdot\left(\boldsymbol{r}_i-\boldsymbol{r}_j\right)}{\left\|\boldsymbol{r}_i-\boldsymbol{r}_j\right\|^3}\dfrac{\partial S_i}{\partial R_{m,x^k}} + \\
\dfrac{\dfrac{\partial \boldsymbol{n}_i}{\partial R_{m,x^k}}_i\cdot\left(\boldsymbol{r}_i-\boldsymbol{r}_j\right)}{\left\|\boldsymbol{r}_i-\boldsymbol{r}_j\right\|^3}S_i + \\
\dfrac{\boldsymbol{n}_i\cdot\left(\dfrac{\partial \boldsymbol{r}_i}{\partial R_{m,x^k}}-\dfrac{\partial \boldsymbol{r}_j}{\partial R_{m,x^k}}\right)}{\left\|\boldsymbol{r}_i-\boldsymbol{r}_j\right\|^3}S_i - \\
3\dfrac{\boldsymbol{n}_i\cdot\left(\boldsymbol{r}_i-\boldsymbol{r}_j\right)}{\left\|\boldsymbol{r}_i-\boldsymbol{r}_j\right\|^5}\left(\left(\boldsymbol{r}_i-\boldsymbol{r}_j\right)\cdot\left(\dfrac{\partial \boldsymbol{r}_i}{\partial R_{m,x^k}}-\dfrac{\partial \boldsymbol{r}_j}{\partial R_{m,x^k}}\right)\right)S_i
\end{array}\right) & \text{for} \quad \mathrm{i} \neq \mathrm{j}, \\
\dfrac{\partial a_{jj}}{\partial R_{m,x^k}} = -\sum_{i\neq j}\dfrac{\partial a_{ij}}{\partial R_{m,x^k}} & \text{for} \quad \mathrm{i} = \mathrm{j}.
\end{cases}
\tag{14}
$$

From this expression it follows that

$$
\sum_{i=1}^{M}\frac{\partial a_{ij}}{\partial R_{m,x^k}} = 0 \quad . \tag{15}
$$

The derivative of the elements of the matrix $B$ over the coordinate of one of the atoms $R_{m,x^k}$ is found as

$$
\frac{\partial b_{ij}}{\partial R_{m,x^k}} = \frac{1}{4\pi}\frac{1-\varepsilon}{1+\varepsilon}\frac{1}{\varepsilon_{in}}\left(
\begin{array}{l}
\dfrac{\boldsymbol{n}_i\cdot\left(\boldsymbol{r}_i-\boldsymbol{R}_j\right)}{\left\|\boldsymbol{r}_i-\boldsymbol{R}_j\right\|^3}\dfrac{\partial S_i}{\partial R_{m,x^k}} + \dfrac{\dfrac{\partial \boldsymbol{n}_i}{\partial R_{m,x^k}}\cdot\left(\boldsymbol{r}_i-\boldsymbol{R}_j\right)}{\left\|\boldsymbol{r}_i-\boldsymbol{R}_j\right\|^3}S_i + \\
\dfrac{\boldsymbol{n}_i\cdot\left(\dfrac{\partial \boldsymbol{r}_i}{\partial R_{m,x^k}}-\delta_{jm}\boldsymbol{e}^k\right)}{\left\|\boldsymbol{r}_i-\boldsymbol{R}_j\right\|^3}S_i - \\
3\dfrac{\boldsymbol{n}_i\cdot\left(\boldsymbol{r}_i-\boldsymbol{R}_j\right)}{\left\|\boldsymbol{r}_i-\boldsymbol{R}_j\right\|^5}\left(\left(\boldsymbol{r}_i-\boldsymbol{R}_j\right)\cdot\left(\dfrac{\partial \boldsymbol{r}_i}{\partial R_{m,x^k}}-\delta_{jm}\boldsymbol{e}^k\right)\right)S_i
\end{array}\right) . \tag{16}
$$

From Eq. (8) it follows that

$$
\sum_{i=1}^{M}\frac{\partial b_{ij}}{\partial R_{m,x^k}} = 0 \; . \tag{17}
$$

The derivative of the elements of the matrix $D$ is calculated as

$$\frac{\partial d_{ij}}{\partial R_{m,x^k}} = -\frac{1}{2}\frac{(\boldsymbol{R}_i - \boldsymbol{r}_j)\cdot\left(\delta_{im}\boldsymbol{e}^k - \frac{\partial \boldsymbol{r}_j}{\partial R_{m,x^k}}\right)}{\left\|\boldsymbol{R}_i - \boldsymbol{r}_j\right\|^3}. \tag{18}$$

### 2.1.1.3. Analytical Gradients of the Energy in the Polarized Continuum Model

Let us introduce the concept of surface elements' mirror charges, which are the solution of the following equation

$$A^T q^* = D^T Q. \tag{19}$$

The physical meaning of the mirror charges $q^*$ consists in the following.

Let there exist the distribution of charges inside a cavity, and these charges produce the surface charge $q_{surf} = BQ$ and the surface element potentials $q^*$ that are found from Eq. (19). Then this system of charges has the same energy as the real surface charges $q$ and the real potentials of surface elements $D^T Q$ . The use of mirror charges allows us to avoid performing the inversion of the matrix when calculating the analytical gradients of the energy.

**The energy can be expressed through the parameters and matrices in several ways.**

$$\Delta G_{pol} = Q^T D q = Q^T D A^{-1} B Q = \left(q^*\right)^T B Q = Q^T B^T q^* = \left(q^*\right)^T A q . \tag{20}$$

The derivative of the energy over the shift of one of the atoms $R_{m,x^k}$ is expressed as

$$\frac{\partial \Delta G_{pol}}{\partial R_{m,x^k}} = \frac{\partial (Q^T D A^{-1} B Q)}{\partial R_{m,x^k}} == Q^T \frac{\partial D}{\partial R_{m,x^k}} q - (q^*)^T \frac{\partial A}{\partial R_{m,x^k}} q + (q^*)^T \frac{\partial B}{\partial R_{m,x^k}} Q . \tag{21}$$

From Eqs. (15), (17), and (21) it follows that the change of all the components of the column $q^*$ by a constant value does not change the energy derivative. Hence, setting the average value of the mirror charge to zero

$$q_i^* = q_i^* - \frac{\sum_{i=1}^{M} q_i^*}{M} \tag{22}$$

does not produce any effect on the energy, but allows us to reduce the numerical and discretization errors.

### 2.1.2. Conductor-Like Screening Model

### 2.1.2.1. Principal Equations for the Energy in the Conductor-Like Screening Model

The conductor-like screening model is used for ε >> 1 [12]. In this model, the surface charge is found at an infinite permittivity ε = ∞. This case is equivalent to the replacement of a surrounding dielectric by a conductor. The advantage of COSMO over PCM consists in multiplying the column of surface charges in the linear matrix equation by a symmetric positive-definite matrix, whose elements do not depend on the normals of surface elements. For such a matrix, the energy and analytical gradients may be found more quickly and precisely.

The COSMO equation in the integral form is written as

$$\int_{\Xi} \frac{\sigma(\boldsymbol{r}')}{|\boldsymbol{r}-\boldsymbol{r}'|} dS + \sum_{i=1}^{N} \frac{Q_i / \varepsilon_{in}}{|\boldsymbol{r}-\boldsymbol{R}_i|} = 0 . \qquad (23)$$

After the discretization of a surface, the COSMO equation is written in the matrix form as

$$A^C q = -\frac{2}{\varepsilon_{in}} D^T Q , \qquad (24)$$

where the matrix $A^C$ is symmetric,

$$(A^C)^T = A^C , \qquad (25)$$

and its element are determined as

$$\begin{cases} a^C{}_{ij} = \left( \dfrac{1}{\|\boldsymbol{r}_i - \boldsymbol{r}_j\|} \right) & i \neq j, \\ a^C{}_{ii} = C_s \dfrac{1}{\sqrt{S_i}} \quad C_S = 2\sqrt{3.83} , & i = j. \end{cases} \qquad (26)$$

The matrix $D$ was determined above in the PCM method.

For COSMO, from the Ostrogradskii–Gauss theorem of electrostatics, we derive the following identity that relates the total surface charge with the total charge inside a cavity

$$\sum_{j=1}^{M} q_j = -\frac{1}{\varepsilon_{in}} \sum_{i=1}^{N} Q_i . \qquad (27)$$

The polar energy for a metal surface ($\frac{1}{2}\sum_i Q_i \int_{\Xi} \frac{\sigma(\boldsymbol{r})}{|\boldsymbol{R}_i - \boldsymbol{r}|} dS`$) is multiplied by the correcting factor $C_f$

$$C_f = \left( \frac{\varepsilon - 1}{\varepsilon + \frac{1}{2}} \right). \tag{28}$$

As a result, the polar solvation energy is determined as

$$\Delta G_{pol} = \frac{1}{2} C_f \sum_i Q_i \int_{\Xi} \frac{\sigma(\boldsymbol{r})}{|\boldsymbol{R}_i - \boldsymbol{r}|} dS` . \tag{29}$$

The relative error in the found energy has a value of the order of 1/(2ε).

The energy in the matrix form is written as

$$\Delta G_{pol} = C_f Q^T D q = -C_f 2 Q^T D \left(A^C\right)^{-1} D^T Q = -\frac{C_f}{2} q^T \left(A^C\right) q. \tag{30}$$

## 2.1.2.2. Analytical Derivatives for the Matrices in the Conductor-Like Screening Model

The derivative of the matrix $A^C$ is written as

$$\begin{cases} \dfrac{\partial a_{ij}^C}{\partial R_{m,x^k}} = -\dfrac{(\boldsymbol{r}_i - \boldsymbol{r}_j)(\dfrac{\partial \boldsymbol{r}_i}{\partial R_{m,x^k}} - \dfrac{\partial \boldsymbol{r}_j}{\partial R_{m,x^k}})}{\left\| \boldsymbol{r}_i - \boldsymbol{r}_j \right\|^3} & \text{for} \quad i \neq j, \\ \dfrac{\partial a_{ii}^C}{\partial R_{m,x^k}} = -2\sqrt{3.83}\, \dfrac{\dfrac{\partial S_i}{\partial R_{m,x^k}}}{2\sqrt{S_i^3}} = -a_{ii}^C \dfrac{\dfrac{\partial S_i}{\partial R_{m,x^k}}}{2 S_i} & \text{for} \quad i = j. \end{cases} \tag{31}$$

The matrix $D$ and its derivative were found above in the part devoted to PCM.

## 2.1.2.3. Analytical Derivatives of the Energy in the Conductor-Like Screening Model

Energy derivative (30) over the coordinate of one of the atoms $R_{m,x^k}$ is calculated as

$$\frac{1}{C_f} \frac{\partial \Delta G_{pol}}{\partial R_{m,x^k}} = -2 \frac{\partial (Q^T D \left(A^C\right)^{-1} D^T Q)}{\partial R_{m,x^k}} = 2 q^T \frac{\partial D^T}{\partial R_{m,x^k}} Q + \frac{1}{2} q^T \frac{\partial A^C}{\partial R_{m,x^k}} q . \tag{32}$$

### 2.1.3. SGB Model

#### 2.1.3.1. Principal Equations for the Energy in the SGB Model

The expression for the electrostatic interaction within the framework of the empirical SGB model [21] is found directly through the parameters of charges inside a cavity and the so-called Born atomic radii $a_i$ as

$$\Delta G_{pol} = -\frac{1}{2}\cdot\frac{1}{\varepsilon_{in}}\frac{1}{1+\frac{1}{2\varepsilon}}\left(1-\frac{1}{\varepsilon}\right)\cdot\sum_{i,j}\frac{Q_i\cdot Q_j}{\sqrt{\left|\boldsymbol{R}_{i,j}\right|^2 + a_i\cdot a_j\cdot exp(-\frac{\left|\boldsymbol{R}_{i,j}\right|^2}{c\cdot a_i\cdot a_j})}}, \qquad (33)$$

where $\boldsymbol{R}_{i,j} = \boldsymbol{R}_i - \boldsymbol{R}_j$, and $c$ is the empirical constant equal to 8.

The Born radii of the atom $i$ are found through the integrals over the surface of a cavity as

$$a_i = \frac{1}{2\left(\sum_n A_n\cdot I_n^i - A_0\right)}, \qquad (34)$$

where $A_n$ are the found empirical constants and $I_n$ is the integrals determined over the surface of a cavity in the form

$$I_n^i = \left[\oint\frac{(\boldsymbol{n}_s\cdot(\boldsymbol{r}_s-\boldsymbol{R}_i))dS}{\left|\boldsymbol{r}_s-\boldsymbol{R}_i\right|^n}\right]^{1/n-3} \quad n\geq 4, \qquad (35)$$

or, through the small surface elements $j$, in the form

$$I_n^i = \left[\sum_j\frac{(\boldsymbol{n}_j\cdot(\boldsymbol{r}_j-\boldsymbol{R}_i))S_j}{\left|\boldsymbol{r}_j-\boldsymbol{R}_i\right|^n}\right]^{1/n-3} \quad n\geq 4. \qquad (36)$$

We restricted our consideration to four integrals with $n$ = 4, 5, 6, 7. The empirically fitted parameters (for the distance expressed in Å) are $A_0$ = –8.191 1/Å, $A_1$ = 152.661, $A_2$ = –238.555, $A_3$ = 151.700, and $A_4$ = –1.074.

#### 2.1.3.2. Analytical Gradients of the Energy in the SGB Model

The total differential of the energy at invariable charges and dielectric constants is found as

$$\frac{\partial \Delta G_{pol}}{\partial R_{m,x^k}} = -\frac{1}{2}\cdot\frac{1}{\varepsilon_{in}}\frac{1}{1+\frac{1}{2\varepsilon}}\left(1-\frac{1}{\varepsilon}\right)\cdot\sum_{i,j} Q_i \cdot Q_j \cdot \frac{\partial g_{i,j}}{\partial R_{m,x^k}}, \tag{37}$$

where the function $g_{i,j}$ and its derivatives are determined as

$$g_{i,j} = \frac{1}{\sqrt{\left|\vec{R}_{i,j}\right|^2 + a_i \cdot a_j \cdot exp(-\frac{|\boldsymbol{R}_{i,j}|^2}{c\cdot a_i \cdot a_j})}}, \tag{38}$$

$$\frac{\partial g_{i,j}}{\partial R_{m,x^k}} = \frac{-g_{i,j}^{\;3}}{2}\left[\begin{array}{l} 2((R_{i,x^k} - R_{j,x^k})\cdot(\delta_{im} - \delta_{jm})\left(1 - \frac{1}{c}exp(-\frac{|\boldsymbol{R}_{i,j}|^2}{c\cdot a_i \cdot a_j})\right) + \\ \left(\frac{\partial a_i}{\partial R_{m,x^k}} a_j + a_i \frac{\partial a_j}{\partial R_{m,x^k}}\right)\left(1 + \frac{|\boldsymbol{R}_{i,j}|^2}{c\cdot a_i \cdot a_j}\right) exp(-\frac{|\boldsymbol{R}_{i,j}|^2}{c\cdot a_i \cdot a_j}) \end{array}\right]. \tag{39}$$

The differentials for the Born atomic radii are calculated as

$$\frac{\partial a_i}{\partial R_{m,x^k}} = \frac{-1}{2\left(\sum_n A_n \cdot I_n^i - A_0\right)^2} \cdot \sum_n A_n \cdot \frac{\partial I_n^i}{\partial R_{m,x^k}}. \tag{40}$$

The derivatives of the integrals $I_n^i$ (for the discrete approximation of integrals, $s$ the index of a surface element) are found as

$$\frac{\partial I_n^i}{\partial R_{m,x^k}} = \frac{\sum_s \frac{\partial J_{s,n}^i}{\partial R_{m,x^k}}}{n-3} \frac{1}{\left[I_n^i\right]^{n-4}} \quad n \geq 4, \tag{41}$$

where $J_{s,n}^i$ and their derivatives are found by the following formulas

$$J_{s,n}^i = \frac{(\boldsymbol{n}_s \cdot (\boldsymbol{r}_s - \boldsymbol{R}_i))S_s}{|\boldsymbol{r}_s - \boldsymbol{R}_i|^n}, \tag{42}$$

$$\frac{\partial J^{i}_{s,n}}{\partial R_{m,x^k}} = \left[ \frac{\left[\left(\boldsymbol{n}_s \cdot (\frac{\partial \boldsymbol{r}_s}{\partial R_{m,x^k}} - \delta_{im}\boldsymbol{e}^k)\right) + \left(\frac{\partial \boldsymbol{n}_s}{\partial R_{m,x^k}} \cdot (\boldsymbol{r}_s - \boldsymbol{R}_i)\right)\right] S_s + \left(\boldsymbol{n}_s \cdot (\boldsymbol{r}_s - \boldsymbol{R}_i)\right)\frac{\partial S_s}{\partial R_{m,x^k}}}{|\boldsymbol{r}_s - \boldsymbol{R}_i|^n} - \frac{n\left(\boldsymbol{n}_s \cdot (\boldsymbol{r}_s - \boldsymbol{R}_i)\right)\left((\frac{\partial \boldsymbol{r}_s}{\partial R_{m,x^k}} - \delta_{im}\boldsymbol{e}^k) \cdot (\boldsymbol{r}_s - \boldsymbol{R}_i)\right) S_s}{|\boldsymbol{r}_s - \boldsymbol{R}_i|^{n+2}} \right]. \tag{43}$$

## 2.2. Non-Polar Component of the Solvation Energy

### 2.2.1. Principal Equations for the Non-Polar Energy Component

To calculate the non-polar energy component, we take SAS elements and use the Bordner–Cavasotto–Abagyan equation [20]

$$E_{SAS}=\sum_j \sigma_j S_j + b, \tag{44}$$

where $j=1,\dots, N_s$ for all the SAS elements with a non-zero surface area.

For water,

$$\sigma_j = \sigma = 0.00378 \text{ kcal/(mol Å}^2\text{) and } b = 0.698 \text{ kcal/mol} \tag{45}$$

### 2.2.2. Gradients of the Non-Polar Component of the Solvation Energy

The gradient of the non-polar energy component is determined as

$$\frac{\partial \Delta G_{nonpol}}{\partial R_{m,x^k}} = \sum_{j=1}^{N} \sigma_j \frac{\partial s_j}{\partial R_{m,x^k}}. \tag{46}$$

The index $j=1,\dots, N_g$ runs over all the SAS elements with a non-zero gradient.

## 3. Computer-Aided Verification of Calculations

The computer-aided verification of the above derived formulas was performed using the DISOLV software [1–3, 7–8] as follows.

### 3.1. Comparing the Numerical and Analytical Gradients

**for the PCM, COSMO, and SGB Methods (Fig. 2)**

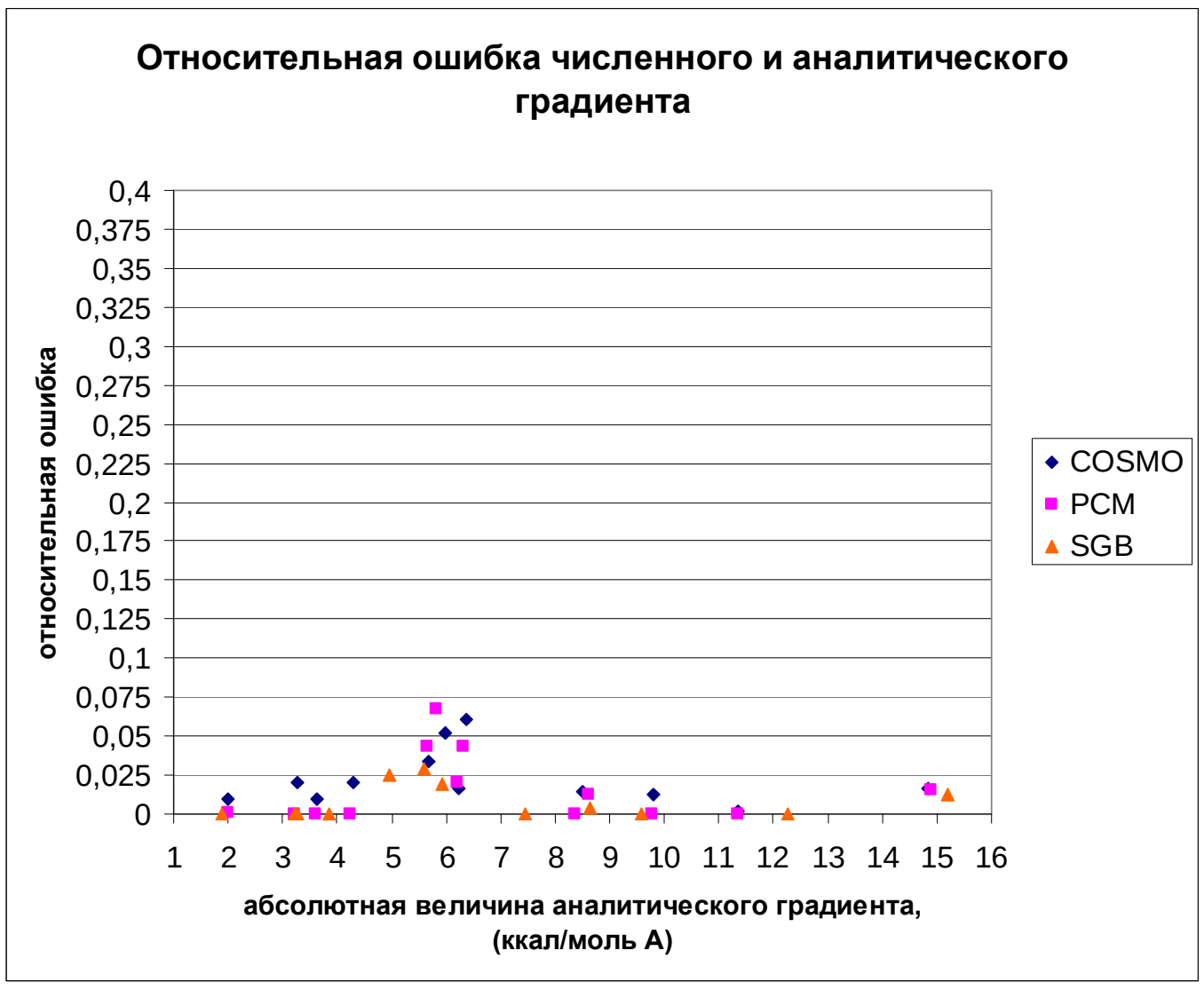

Fig. 2. Relative error in the numerical and analytical gradients for the PCM, COSMO, and SGB methods (at a grid step of 0.1). The numerical gradient (COSMO or SGB) is plotted along the *x* axis, and the relative error in the energy with respect to the analytical gradient (COSMO or SGB, respectively) is plotted along the *y* axis,

$$\delta^{COSMO} = \frac{|\nabla_{\boldsymbol{R}_j} E_{num} - \nabla_{\boldsymbol{R}_j} E_{anal}|}{|\nabla_{\boldsymbol{R}_j} E_{num}|}.$$

Key:

Относительная ошибка численного и аналитического градиента --> Relative error in the numerical and analytical gradient;

относительная ошибка --> relative error;

абсолютная величина аналитического градиента, (ккал/моль Å) --> absolute value of the analytical gradient, (kcal/(mol Å))

For the PCM, COSMO, and SGB methods, we have obtained the following results on the comparison of numerical and analytical gradients:

(1) For PCM, the relative error in the numerical derivative with respect to the analytical derivative for gradients exceeding 2 kcal/(mol Å) is less than 6.6 % and, on the average, equals 2%. For volumetric atoms, the error in the numerical derivative with respect to the analytical derivative for gradients exceeding 2 kcal/(mol Å) is less than 0.03%.

(2) For SGB, the relative error in the numerical derivative with respect to the analytical derivative for gradients exceeding 2 kcal/(mol Å) is less than 3% and, on the average, equals 1.5%. For volumetric atoms, the error in the numerical derivative with respect to the analytical derivative for gradients exceeding 2 kcal/(mol Å) is less than 0.03%.

(3) For COSMO, the relative error in the numerical derivative with respect to the analytical derivative for gradients exceeding 2 kcal/(mol Å) is less than 6% and, on the average, equals 3%. For volumetric atoms, the error in the numerical derivative with respect to the analytical derivative for gradients exceeding 2 kcal/(mol Å) is less than 0.2%.

The errors in the numerical gradients of volumetric atoms are determined by the error in calculating the energy, and the errors in the numerical gradients of surface atoms are determined first of all by the shift of the grid upon the shift of an atom.

### 3.2. Comparing the Analytical Gradients for the COSMO and SGB Methods with the Analytical Gradients for the PCM Method

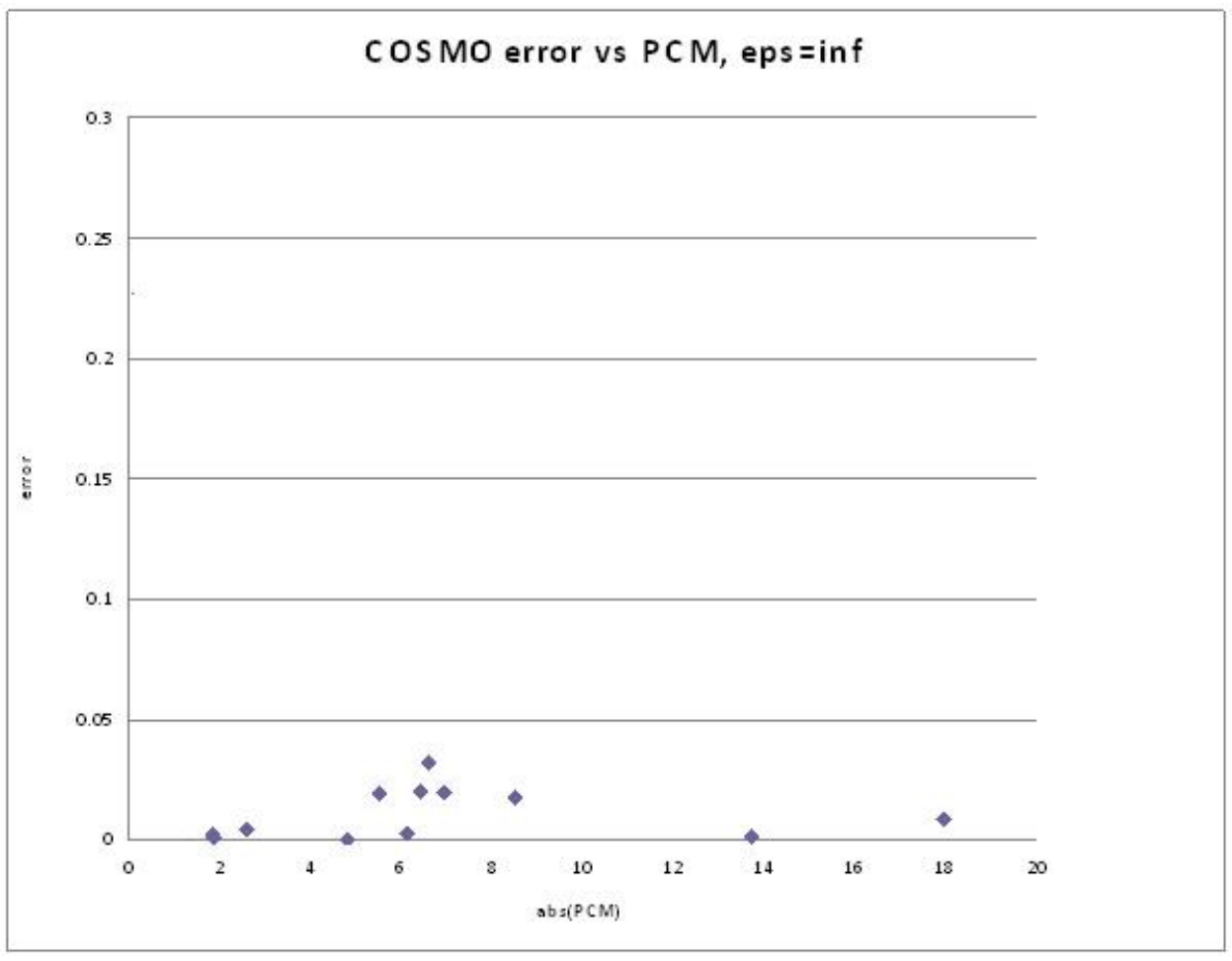

Fig. 3. Relative error in the analytical gradient for the COSMO method in comparison with the PCM method at $\varepsilon \to \infty$. The PCM gradient is plotted along the $x$ axis, and the relative error with respect to the COSMO method is plotted along the $y$ axis,

$$\delta = \frac{|\nabla_{R_j} E_{COSMO} - \nabla_{R_j} E_{PCM}|}{|\nabla_{R_j} E_{PCM}|}$$

(1) Let us compare the analytical gradients for the COSMO and PCM methods (Fig. 3).

(a) For the dielectric permittivity tending to infinity ε→∞ and the grid step tending to zero, the analytical gradients for the COSMO and PCM method must precisely coinside (since COSMO is the limit of PCM at an infinite dielectric permittivity and an infinitely small grid step). Really, at ε→∞, the relative error in the COSMO analytical derivative with espect to the PCM analytical derivative for gradients exceeding 2 kcal/(mol Å) is less than 2.5% and, on the average, equals 1.2%. Maximum values of the relative error in the analytical derivative correspond to surface atoms;

(b) At ε = 78.5 (water), the relative error in the COSMO analytical derivative with respect to the PCM analytical derivative is slightly higher, namely, for gradients exceeding 2 kcal/(mol Å) is less than 4% and, on the average, equals 2%.

(2) Let us compare the analytical gradients for the SGB and PCM models (Fig. 4).

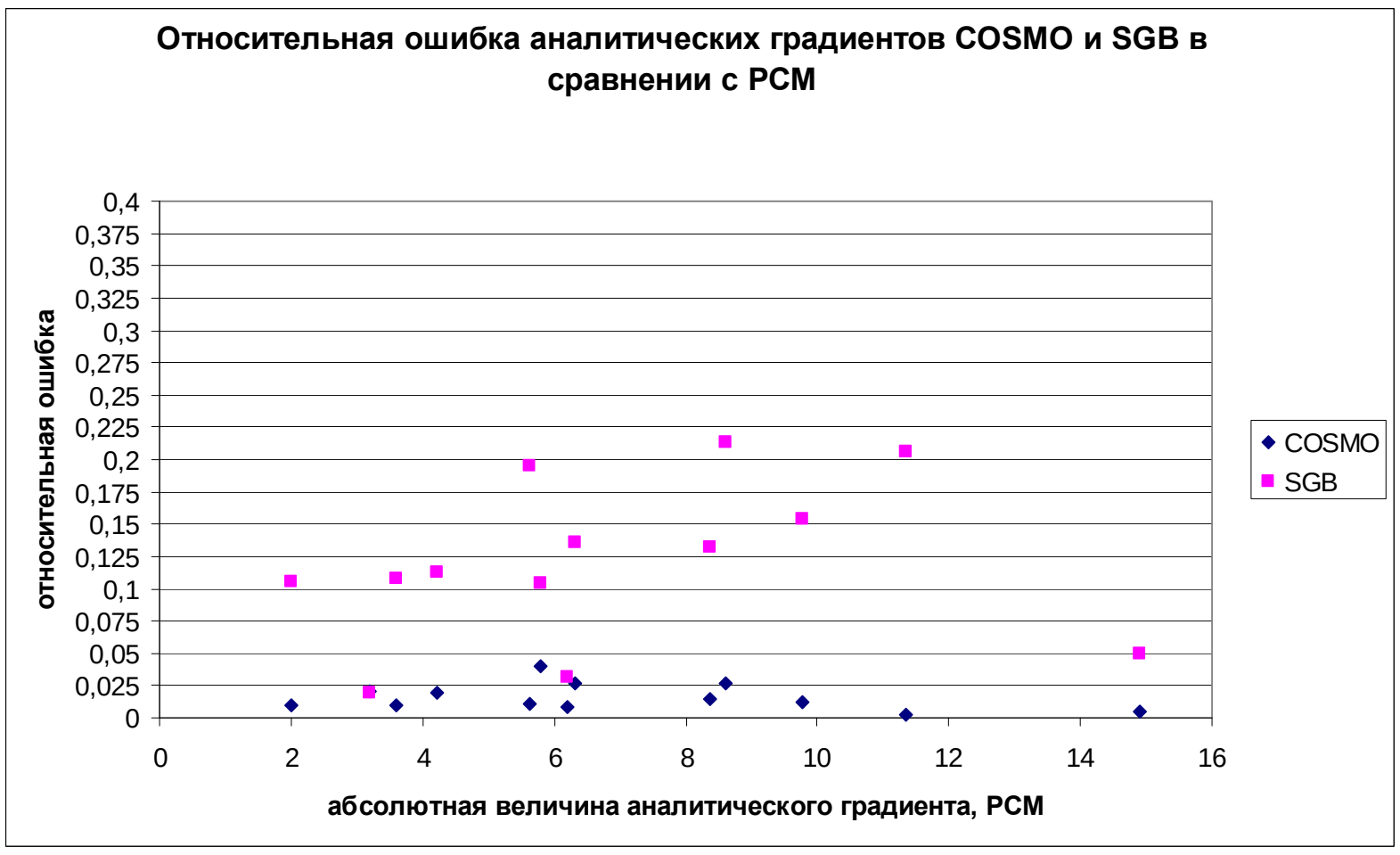

Fig. 4. Relative error in the analytical gradients for the COSMO and SGB methods in comparison with the PCM method at $\varepsilon$ = 78.5. The PCM analytical gradient is plotted

along the $x$ axis, and the relative error with respect to the COSMO or SGB methods is plotted along the $y$ axis, $\delta = \frac{|\nabla_{R_j} E_{(COSMO,SGB)} - \nabla_{R_j} E_{PCM}|}{|\nabla_{R_j} E_{PCM}|}$.

Key:

Относительная ошибка аналитических градиентов COSMO и SGB в сравнении с PCM --> Relative error in the analytical gradients for the COSMO and SGB methods in comparison with the PCM method;

относительная ошибка --> relative error;

абсолютная величина аналитического градиента PCM --> absolute value of the PCM analytical gradient

At ε = 78.5 (water), the relative error in the SGB analytical derivative with respect to the PCM analytical derivative is much higher, namely, for gradients exceeding 2 kcal/(mol Å) is less than 23% and, on the average, equals 13%. It is not surprising, as SGB is a rough empirical model.

## Conclusions

In this work, the analytical gradients of the energy were calculated on the basis of the PCM, COSMO, and SGB methods for a molecular surface obtained via primary and secondary rolling by the algorithms described in [1–9].

Let us list the problem cases, which may arise when calculating the gradients and solving the problem of optimizing the configuration of molecules with the use of these gradients:

- Degeneration, when a triple point is supported by several atoms instead of three atoms (e.g., aromatic rings);
- The sharp reconstruction of a molecular surface due to that the secondary rolling radius or the critical distance are changed by the program itself in the case of their automatic adjustment;

- The sharp reconstruction of a molecular surface due to the appearance or disappearance of a rolling torus (primary or secondary) upon the shift of triple points or the change of a narrow torus neck at distances close to the critical distance; and
- The sharp reconstruction of a molecular surface due to the appearance or disappearance of primary or secondary triple points.

As a result, there may be gradient jumps and optimizer's "oscillations" near the above described points.

To confirm the correctness of the formulas for the energy gradients obtained by different methods, we performed computer-aided calculations. Good agreement of the analytical gradients obtained by different methods with each other and good agreement of the numerical and analytical gradients for the COSMO method confirm the correctness of the derived formulas.

## Acknowledgments

The given study is a more detailed presentation and further development of the work [9]. We are profoundly grateful to all the authors for their studies taken as the basis for the given paper.

УДК 535.71

# КОНТИНУАЛЬНАЯ МОДЕЛЬ СРЕДЫ IV: ВЫЧИСЛЕНИЕ АНАЛИТИЧЕСКИХ ГРАДИЕНТОВ ДЛЯ ЭНЕРГИИ СОЛЬВАТАЦИИ ПО КООРДИНАТАМ АТОМОВ.

Купервассер* О.Ю., Ваннер** Н.Э.
*ООО «Транзист Видео», участник Сколково
***Государственное научное учреждение Всероссийский научно-исследовательский институт ветеринарной санитарии, гигиены и экологии Россельхозакадемии, Москва*
*E-mail: olegkup@yahoo.com

В данной работе мы описываем методы нахождение аналитических градиентов (производных) по координатам атомов от энергии сольватации. Это делается как для неполярной составляющей энергии, так и для полярной составляющей энергии, полученной методами PCM, COSMO и SGB. Эти градиенты ищутся с использованием аналитических градиентов параметров (координат, нормалей и площадей) поверхностных элементов. Поверхностные элементы расположены как на оптимально гладкой молекулярной поверхности исключенного объема (SES), так и производной от нее поверхности доступной растворителю (SAS). Эти поверхности находится методами первичной и вторичной обкатки, согласно алгоритму, описанному в [1-9].

Ключевые слова: энергия сольватации, аналитические градиенты, численные градиенты, молекулярная поверхность, первичная обкатка, вторичная обкатка

## 1. Введение

Свободная энергия сольватации $\Delta G_s$ в континуальных моделях растворителя представляется в виде суммы трех составляющих [10]

$$\Delta G_s = \Delta G_{cav} + \Delta G_{np} + \Delta G_{pol} ,$$

(1)

$$\Delta G_{nonpol} = \Delta G_{cav} + \Delta G_{np} ,$$

где

$\Delta G_{pol}$ –полярная составляющая, связанная с поляризацией диэлектрика-раствора

$\Delta G_{cav}$ – кавитационная составляющая, связанная с образованием полости в растворителе

$\Delta G_{np}$ - составляющая, определяемая ванн-дер-ваальсовым взаимодействием

$\Delta G_{pol}$ – неполярная составляющая, являющаяся суммой предыдущих двух.

Приведем краткий обзор работ, посвященных нахождению аналитических градиентов (производных) энергии сольватации, которые берутся по координатам атомов. Подобные градиенты необходимы, например, при финальной оптимизации положения и геометрии молекулы - лиганда в модели доркинга.

Задача нахождения аналитических градиентов распадается на две подзадачи:

1) Нахождение аналитических градиентов параметров (координат, нормали, площади) поверхностных элементов молекулярной поверхности.

Эта задача для алгоритма GEPOL построения поверхности (путем заполнения пустого пространства фиктивными шарами) рассмотрена в [11]. Алгоритм, реализующей построение гладкой поверхности методами первичной и вторичной обкатки, описан в [1-9].

2) На основе полученных выше градиентов параметров поверхностных элементов рассчитываются аналитические градиенты матриц, входящих в уравнения модели (для COSMO [12-13], для PCM [14-15]) .

На основе уже градиентов этих матриц затем рассчитываются и аналитические градиенты энергии сольватации.

Следует отметить, что для метода PCM описанная в [14-15] методика нахождения градиентов требует обращения одной из матриц уравнения PCM. Если же число поверхностных элементов велико, то и размер этой матрицы велик. Хранение такой матрицы требует большой компьютерной памяти и значительно повышает вычислительное время задачи. Кроме того сами алгоритмы обращения матриц [16-17] также требуют много памяти и затрат времени. Однако для метода с укрупненными поверхностными элементами [18] обратная матрица мала, ее вычисление и использование не проблематично. Используя методику в [19] можно итерационно найти обратную матрицу. Используя методы, изложенные в [14-15], можно уже найти и сами аналитические градиенты.

В данной работе мы описываем решение второй подзадачи – нахождение аналитических градиентов для полярной составляющей энергии сольватации для методов PCM [20], COSMO [12] и SGB [21] , а также для неполярной составляющей энергии сольватации [20]. Для этого используются аналитические градиенты параметров поверхностных элементов молекулярных поверхностей, полученных методами первичной и вторичной обкатки [1-9]

Существуют два вида поверхности вокруг молекулы [1-9]. Во-первых, это SES (Solvent Excluded Surface) - поверхность исключённого из растворителя объёма. Объем, занимаемый растворителем, лежит *вне* объема, ограниченного этой поверхностью. Сам субстрат полностью лежит *внутри* этого объема. Алгоритм [1-9] методами первичной и вторичной стоит гладкую поверхность SES. Во-вторых, это SAS (Solvent Accessible Surface) - поверхность доступная растворителю, которая образуется центрами молекул растворителя, касающихся молекулы субстрата при первичной обкатке. Первый тип поверхности используется для расчета полярной составляющей энергии сольватации (действительно, на SAS многие участки молекулы, существенные для электростатического взаимодействия, имеют нулевую или очень малую площадь [22]), а второй – для расчета кавитационной о ванн-дер-ваальсовской составляющих. Расчеты аналитических градиентов энергии, полученной методами PCM, COSMO и SGB реализованы в программе DISOLV [1-9]

## 2. Физические модели

### 2.1 Полярная часть энергии сольватации.

Для расчета неполяной части энергии берутся поверхностные элементы на SES.

#### 2.1.1 Модель PCM.

##### 2.1.1.1 Основные уравнения для энергии PCM.

Дадим теперь точное описание используемых нами моделей.

Начнем с модели *PCM (Polarized Continuum Model)* [16], [19-20]. Рассмотрим описанную выше задачу нахождения полярной части энергии сольватации. При этом раствор $\Omega_{ex}$ заменяется диэлектриком с известной диэлектрической проницаемостью $\varepsilon_{ex}$. Растворенная молекула рассматривается как полость $\Omega_{in}$ внутри этого диэлектрика, заполненная либо вакуумом $\varepsilon_{in}$=1, либо диэлектриком с диэлектрической проницаемостью $\varepsilon_{in}\neq 1$. Внутри полости находятся точечные заряды $Q_i$, находящихся в центрах атомов $\boldsymbol{R}_i$ на некотором удалении от границы $\Xi$. Эти заряды, приближенно аппроксимирующие распределение заряда внутри молекулы, мы находили методами силового поля MMFF94 [23-25]. Возможна ситуация, когда имеется не одна, а несколько полостей, соответствующих нескольким молекулам. При этом влияние диэлектрика-раствора полностью определяется поверхностным зарядом $\sigma(\boldsymbol{r})$ на границе полости. Точное линейное уравнение, связывающее этот поверхностный заряд с положением и величиной зарядов внутри поверхности, назовется PCM. Для нахождения полярной компоненты энергии сольватации мы вначале считаем энергию $E_1$, необходимую для перенесения молекулы из диэлектрика $\varepsilon_{in}$ в вакуум (поверхностный заряд, который при этом образуется, имеет плотность $\sigma_{vac}(\boldsymbol{r})$). Затем энергию $E_2$, необходимую для перенесения молекулы из диэлектрика $\varepsilon_{in}$ в диэлектрик $\varepsilon_{ex}$. Энергия сольватации перехода одного моля субстрата из вакуума в растворитель $\varepsilon_{ex}$ равна разнице этих двух энергий: $\Delta G_{pol} = E_2 - E_1$ и описывается следующей формулой:

$$\Delta G_{pol} = \frac{1}{2}\sum_i Q_i \int_{\Xi} \frac{\sigma(\boldsymbol{r}) - \sigma_{vac}(\boldsymbol{r})}{|\boldsymbol{R}_i - \boldsymbol{r}|} dS` .$$

(2)

Для $\varepsilon_{in}=1$ поверхностный заряд в вакууме $\sigma_{vac}=0$ и, следовательно

$$\Delta G_{pol} = \frac{1}{2}\sum_i Q_i \int_{\Xi} \frac{\sigma(\boldsymbol{r})}{|\boldsymbol{R}_i - \boldsymbol{r}|} dS` .$$

(3)

Поскольку $\varepsilon_{in}$ это эмпирически подбираемый параметр, мы выбирали его в расчетах равным $\varepsilon_{in}=1$, чтобы исключить из рассмотрения вакуумную составляющую энергии и упростить расчеты.

Интегральное уравнение PCM для поверхностного заряда имеет следующий вид:

$$\sigma(\boldsymbol{r}) = \frac{\varepsilon_{in} - \varepsilon_{ex}}{2\pi(\varepsilon_{in} + \varepsilon_{ex})}\left(\sum_i \frac{Q_i((\boldsymbol{r} - \boldsymbol{R}_i)\cdot \boldsymbol{n})}{\varepsilon_{in}|\boldsymbol{r} - \boldsymbol{R}_i|^3} + \int_{\Xi} \frac{\sigma(\boldsymbol{r}`)((\boldsymbol{r} - \boldsymbol{r}`)\cdot \boldsymbol{n})}{|\boldsymbol{r} - \boldsymbol{r}`|^3} dS`\right).$$

(4)

Последний интеграл является сингулярным при $\boldsymbol{r}=\boldsymbol{r}`$ и по определению равен:

$$\int_{\Xi} \frac{\sigma(\boldsymbol{r}`)((\boldsymbol{r} - \boldsymbol{r}`)\cdot \boldsymbol{n})}{|\boldsymbol{r} - \boldsymbol{r}`|^3} dS` = \lim_{\delta \to 0} \int_{\Xi / (|\boldsymbol{r} - \boldsymbol{r}`| < \delta)} \frac{\sigma(\boldsymbol{r}`)((\boldsymbol{r} - \boldsymbol{r}`)\cdot \boldsymbol{n})}{|\boldsymbol{r} - \boldsymbol{r}`|^3} dS` . \quad (5)$$

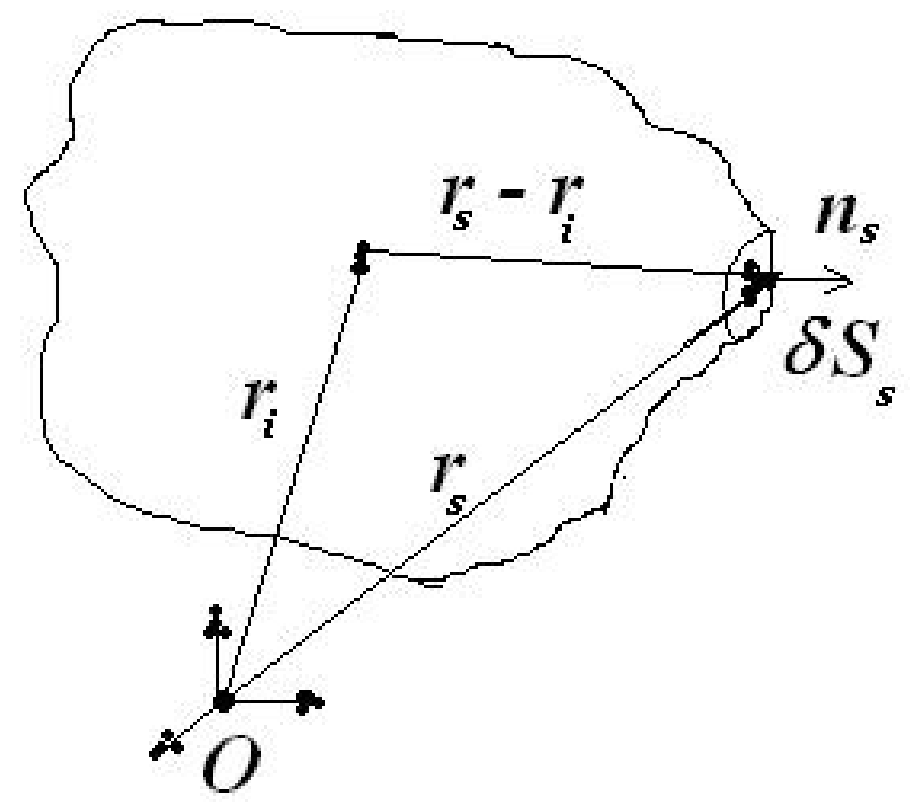

Рис. 1 Элемент поверхности относительно начала координат.

После дискретизации поверхности (Рис.1) (т.е. после разбиения её на малые поверхностные элементы, несущие поверхностный заряд) это линейное уравнение может быть записано в матричной форме:

$$Aq = BQ \,, \quad (6)$$

A – матрица, зависящая от параметров поверхностных элементов

B – матрица, зависящая от геометрических параметров поверхностных элементов и положения зарядов внутри молекулы.

q – столбец зарядов поверхностных элементов

Q - столбец зарядов внутри молекулы.

Определим параметры малых поверхностных элементов и найдем в дальнейшем, как через них выражаются элементы описанных выше матриц:

$M$ - число поверхностных элементов

$N$ - число зарядов внутри полости

$q_j$ - заряд $j$-ого поверхностного элемента.

$S_j$ – площадь поверхностного элемента

$\boldsymbol{r}_j$ –радиус-вектор центра поверхностного элемента

$\boldsymbol{n}_j$ – нормаль в центре поверхностного элемента

$Q_i$ – заряды внутри молекулы

$\boldsymbol{R}_i$ – радиус-вектор заряда i внутри молекулы

$\varepsilon=\varepsilon_{out}/\varepsilon_{in}$

Элементы $b_{ij}$ матрицы $B$:

$$b_{ij} = \frac{\boldsymbol{n}_i\left(\boldsymbol{r}_i - \boldsymbol{R}_j\right)}{\left\|\boldsymbol{r}_i - \boldsymbol{R}_j\right\|^3} S_i \frac{1-\varepsilon}{4\pi(\varepsilon+1)} \frac{1}{\varepsilon_{in}}.$$

(7)

Для столбцов матрицы $B$ выполняется тождество

$$\sum_{i=1}^{M} b_{ij} = \frac{1-\varepsilon}{1+\varepsilon} \frac{1}{\varepsilon_{in}}.$$

(8)

Элементы матрицы $A$:

$$a_{ij} = \begin{cases} \dfrac{\boldsymbol{n}_i\left(\boldsymbol{r}_i - \boldsymbol{r}_j\right)}{\left\|\boldsymbol{r}_i - \boldsymbol{r}_j\right\|^3} S_i \dfrac{\varepsilon - 1}{4\pi(\varepsilon+1)} & i \neq j \\ a_{jj} = \dfrac{\varepsilon}{\varepsilon+1} - \sum\limits_{i \neq j} a_{ij} & i = j \end{cases}.$$

(9)

Для столбцов матрицы $A$ выполняется тождество:

$$\sum_{i=1}^{M} a_{ij} = \frac{\varepsilon}{\varepsilon+1}.$$

(10)

Из (8) и (10) автоматически следует тождество, связывающее суммарный поверхностный заряд с суммарным зарядом внутри полости:

$$\sum_{j=1}^{M} q_j = -\left(\frac{1}{\varepsilon_{in}} - \frac{1}{\varepsilon_{ex}}\right) \sum_{i=1}^{N} Q_i .$$

(11)

Использование (8) для коррекции численных ошибок автоматически приводит к очень точному выполнению указанного выше тождества. Каждая из двух компонент энергии в (2) или энергия (3) могут быть записаны в матричной форме:

$$\Delta G_{pol} = Q^T Dq$$

(12)

Где матричные элементы D:

$$d_{ij} = \frac{1}{2}\left(\frac{1}{\left\| \boldsymbol{R}_i - \boldsymbol{r}_j \right\|}\right) \qquad (13)$$

### 2.1.1.2 Аналитические производные для матриц

$R_{i,x^k}$ - $x^k$-ая компонента радиус-вектора $R_i$

$x^k$ член множества $\{ x^0, x^1, x^2 \}$

$i$ – номер сдвигаемого атома ($0<i<N$);

$\boldsymbol{e}^k$ - базисный орт в направлении координаты $x^k$, член множества $\{ \boldsymbol{e}^0, \boldsymbol{e}^1, \boldsymbol{e}^2 \}$

$\delta_{ij}$ − дельта-функция

Производная элементов матрицы A по координате одного из атомов $R_{m,x^k}$ :

$$
\begin{cases}
\dfrac{\partial a_{ij}}{\partial R_{m,x^k}} = \dfrac{1}{4\pi}\dfrac{\varepsilon-1}{\varepsilon+1}\left(
\begin{array}{l}
\dfrac{\boldsymbol{n}_i\cdot(\boldsymbol{r}_i-\boldsymbol{r}_j)}{\|\boldsymbol{r}_i-\boldsymbol{r}_j\|^3}\dfrac{\partial S_i}{\partial R_{m,x^k}} + \\
\dfrac{\dfrac{\partial \boldsymbol{n}_i}{\partial R_{m,x^k}}{}_i\cdot(\boldsymbol{r}_i-\boldsymbol{r}_j)}{\|\boldsymbol{r}_i-\boldsymbol{r}_j\|^3} S_i + \\
\dfrac{\boldsymbol{n}_i\cdot\left(\dfrac{\partial \boldsymbol{r}_i}{\partial R_{m,x^k}}-\dfrac{\partial \boldsymbol{r}_j}{\partial R_{m,x^k}}\right)}{\|\boldsymbol{r}_i-\boldsymbol{r}_j\|^3} S_i - \\
3\dfrac{\boldsymbol{n}_i\cdot(\boldsymbol{r}_i-\boldsymbol{r}_j)}{\|\boldsymbol{r}_i-\boldsymbol{r}_j\|^5}\left((\boldsymbol{r}_i-\boldsymbol{r}_j)\cdot\left(\dfrac{\partial \boldsymbol{r}_i}{\partial R_{m,x^k}}-\dfrac{\partial \boldsymbol{r}_j}{\partial R_{m,x^k}}\right)\right)S_i
\end{array}\right) & \text{for } i \neq j, \\
\dfrac{\partial a_{jj}}{\partial R_{m,x^k}} = -\sum_{i\neq j}\dfrac{\partial a_{ij}}{\partial R_{m,x^k}} & \text{for } i = j.
\end{cases}
\tag{14}
$$

Отсюда следует, что

$$
\sum_{i=1}^{M}\frac{\partial a_{ij}}{\partial R_{m,x^k}} = 0 \quad .
$$

(15)

Производная элементов матрицы B по координате одного из атомов $R_{m,x^k}$ :

$$\frac{\partial b_{ij}}{\partial R_{m,x^k}} = \frac{1}{4\pi}\frac{1-\varepsilon}{1+\varepsilon}\frac{1}{\varepsilon_{in}}\left( \begin{array}{l} \dfrac{\boldsymbol{n}_i \cdot (\boldsymbol{r}_i - \boldsymbol{R}_j)}{\|\boldsymbol{r}_i - \boldsymbol{R}_j\|^3}\dfrac{\partial S_i}{\partial R_{m,x^k}} + \dfrac{\dfrac{\partial \boldsymbol{n}_i}{\partial R_{m,x^k}} \cdot (\boldsymbol{r}_i - \boldsymbol{R}_j)}{\|\boldsymbol{r}_i - \boldsymbol{R}_j\|^3} S_i + \\ \dfrac{\boldsymbol{n}_i \cdot \left( \dfrac{\partial \boldsymbol{r}_i}{\partial R_{m,x^k}} - \delta_{jm}\boldsymbol{e}^k \right)}{\|\boldsymbol{r}_i - \boldsymbol{R}_j\|^3} S_i - \\ 3\dfrac{\boldsymbol{n}_i \cdot (\boldsymbol{r}_i - \boldsymbol{R}_j)}{\|\boldsymbol{r}_i - \boldsymbol{R}_j\|^5}\left( (\boldsymbol{r}_i - \boldsymbol{R}_j) \cdot \left( \dfrac{\partial \boldsymbol{r}_i}{\partial R_{m,x^k}} - \delta_{jm}\boldsymbol{e}^k \right) \right) S_i \end{array} \right).$$

(16)

Из формул (8) следует, что

$$\sum_{i=1}^{M} \frac{\partial b_{ij}}{\partial R_{m,x^k}} = 0 .$$

(17)

Производная элементов матрицы $D$.

$$\frac{\partial d_{ij}}{\partial R_{m,x^k}} = -\frac{1}{2}\frac{(\boldsymbol{R}_i - \boldsymbol{r}_j) \cdot \left( \delta_{im}\boldsymbol{e}^k - \dfrac{\partial \boldsymbol{r}_j}{\partial R_{m,x^k}} \right)}{\|\boldsymbol{R}_i - \boldsymbol{r}_j\|^3}.$$

(18)

### 2.1.1.3 Аналитические градиенты энергии для PCM.

Введем понятие зеркальных зарядов поверхностных элементов, которые являются решением следующего уравнения:

$$A^T q^* = D^T Q.$$

(19)

Физический смысл зеркальных зарядов $q^*$ следующий:
Пусть имеется распределение зарядов внутри полости, создающие поверхностный заряд $q_{surf} = BQ$ и потенциалы поверхностных элементов $q^*$, которые находятся из (19). Тогда эта система зарядов имеет ту же энергию, что и настоящие поверхностные заряды q и настоящие потенциалы поверхностных элементов $D^T Q$. Использование зеркальных зарядов позволяет избежать операции обращения матрицы при нахождении аналитических градиентов энергии.

**Энергия через параметры и матрицы выражается несколькими способами:**

$$\Delta G_{pol} = Q^T Dq = Q^T DA^{-1}BQ = \left(q^*\right)^T BQ = Q^T B^T q^* = \left(q^*\right)^T Aq.$$

(20)

Производная энергии по смещению одного из атомов $R_{m,x^k}$:

$$\frac{\partial \Delta G_{pol}}{\partial R_{m,x^k}} = \frac{\partial (Q^T DA^{-1}BQ)}{\partial R_{m,x^k}} == Q^T \frac{\partial D}{\partial R_{m,x^k}} q - (q^*)^T \frac{\partial A}{\partial R_{m,x^k}} q + (q^*)^T \frac{\partial B}{\partial R_{m,x^k}} Q.$$

(21)

Из формул (15), (17) и (21) следует, что что изменение всех компонент столбца $q^*$ на постоянную величину не меняет значения производной энергии. Отсюда обнуление среднего значения зеркального заряда

$$q_i^* = q_i^* - \frac{\sum_{i=1}^{M} q_i^*}{M}$$

(22)

не влияет на энергию, но позволяет уменшить численные ошибки и ошибки дискретизациии.

### 2.1.2 Модель COSMO.

#### 2.1.2.1 Основные уравнения для энергии COSMO.

Для случая больших значений ε>>1 используется *модель COSMO ("**CO**nductor-like **S**creening **MO**del")* [12]. В этой модели поверхностный заряд находится для бесконечной проницаемости ε=∞. Этот случай эквивалентен замене окружающего диэлектрика на проводник. Преимущество COSMO над PCM состоит в том, что в линейном матричном уравнении столбец поверхностных зарядов помножается на симметричную, положительно определенную матрицу, элементы которой не зависят от нормалей поверхностных элементов. Для такой матрицы нахождение энергии и аналитических градиентов можно сделать быстрее и точнее.

Уравнение COSMO в интегральной форме записывается в следующем виде:

$$\int_{\Xi} \frac{\sigma(\boldsymbol{r}')}{|\boldsymbol{r}-\boldsymbol{r}'|} dS + \sum_{i=1}^{N} \frac{Q_i / \varepsilon_{in}}{|\boldsymbol{r}-\boldsymbol{R}_i|} = 0 .$$

(23)

Уравнение COSMO в матричном виде после дискретизации поверхности записывается в следующем виде

$$A^C q = -\frac{2}{\varepsilon_{in}} D^T Q ,$$

(24)

где матрица $A^C$ симметрична

$$(A^C)^T = A^C ,$$

(25)

и ее элементы:

$$\begin{cases} a^C{}_{ij} = \left( \frac{1}{\left\| \boldsymbol{r}_i - \boldsymbol{r}_j \right\|} \right) & i \neq j, \\ a^C{}_{ii} = C_s \frac{1}{\sqrt{S_i}} \quad C_S = 2\sqrt{3.83}, & i = j. \end{cases}$$

(26)

Матрица *D* была определена выше в методе PCM.

Для COSMO из теоремы Остроградского-Гаусса электростатики следует тождество, связывающее суммарный поверхностный заряд с суммарным зарядом внутри полости:

$$\sum_{j=1}^{M} q_j = -\frac{1}{\varepsilon_{in}} \sum_{i=1}^{N} Q_i .$$

(27)

Полярная энергия, получаемая для металлической поверхности ($\frac{1}{2}\sum_i Q_i \int_{\Xi} \frac{\sigma(\boldsymbol{r})}{|\boldsymbol{R}_i - \boldsymbol{r}|} dS`$), помножается на корректирующий фактор $C_f$:

$$C_f = \left( \frac{\varepsilon - 1}{\varepsilon + \frac{1}{2}} \right) .$$

(28)

В итоге полярная энергия сольватации:

$$\Delta G_{pol} = \frac{1}{2} C_f \sum_i Q_i \int_{\Xi} \frac{\sigma(\boldsymbol{r})}{|\boldsymbol{R}_i - \boldsymbol{r}|} dS` .$$

(29)

Относительная погрешность найденной энергии имеет величину порядка 1/(2ε).

Энергия в матричной форме:

$$\Delta G_{pol} = C_f Q^T D q = -C_f 2 Q^T D \left(A^C\right)^{-1} D^T Q = -\frac{C_f}{2} q^T \left(A^C\right) q .$$

(30)

#### 2.1.2.2 Аналитические производные для матриц COSMO.

Производная матрицы $A^C$:

$$\begin{cases} \dfrac{\partial a^C_{ij}}{\partial R_{m,x^k}} = -\dfrac{(\boldsymbol{r}_i - \boldsymbol{r}_j)\left(\dfrac{\partial \boldsymbol{r}_i}{\partial R_{m,x^k}} - \dfrac{\partial \boldsymbol{r}_j}{\partial R_{m,x^k}}\right)}{\left\|\boldsymbol{r}_i - \boldsymbol{r}_j\right\|^3} \quad \text{for} \quad i \neq j, \\ \dfrac{\partial a^C_{ii}}{\partial R_{m,x^k}} = -2\sqrt{3.83}\,\dfrac{\dfrac{\partial S_i}{\partial R_{m,x^k}}}{2\sqrt{S_i^3}} = -a^C_{ii}\,\dfrac{\dfrac{\partial S_i}{\partial R_{m,x^k}}}{2S_i} \quad \text{for} \quad i = j. \end{cases}$$

(31)

Матрица $D$ и ее производная найдена ранее в разделе о PCM.

#### 2.1.2.3 Аналитические производные для энергии COSMO.

Производная энергии (30 ) по координате одного из атомов $R_{m,x^k}$ :

$$\frac{1}{C_f}\frac{\partial \Delta G_{pol}}{\partial R_{m,x^k}} = -2\frac{\partial (Q^T D\left(A^C\right)^{-1} D^T Q)}{\partial R_{m,x^k}} = 2q^T\frac{\partial D^T}{\partial R_{m,x^k}}Q + \frac{1}{2}q^T\frac{\partial A^C}{\partial R_{m,x^k}}q\,.$$

(32)

### 2.1.3 Модель SGB.

#### 2.1.3.1 Основные уравнения для энергии SGB.

Выражение для электростатического взаимодействия в рамках эмпирической модели SGB [21] находится напрямую через параметры зарядов внутри полости и так называемые Борновские радиусы атомов $a_i$:

$$\Delta G_{pol} = -\frac{1}{2}\cdot\frac{1}{\varepsilon_{in}}\frac{1}{1+\dfrac{1}{2\varepsilon}}\left(1-\frac{1}{\varepsilon}\right)\cdot\sum_{i,j}\frac{Q_i\cdot Q_j}{\sqrt{\left|\boldsymbol{R}_{i,j}\right|^2 + a_i\cdot a_j\cdot exp(-\frac{\left|\boldsymbol{R}_{i,j}\right|^2}{c\cdot a_i\cdot a_j})}},$$

(33)

Где $\boldsymbol{R}_{i,j} = \boldsymbol{R}_i - \boldsymbol{R}_j$; $c$- эмпирическая константа равная 8.

Борновские радиусы атома i находятся через интегралы по поверхности полости:

$$a_i = \frac{1}{2\left(\sum_n A_n \cdot I_n^i - A_0\right)},$$

(34)

$A_n$ – находимые эмпирически константы, $I_n$ – интегралы по поверхности полости вида:

$$I_n^i = \left[ \oint \frac{(\boldsymbol{n}_s \cdot (\boldsymbol{r}_s - \boldsymbol{R}_i))dS}{|\boldsymbol{r}_s - \boldsymbol{R}_i|^n} \right]^{1/n-3} \quad n \geq 4 \ ,$$

(35)

или, выражая интегралы через малые поверхностные элементы j,

$$I_n^i = \left[ \sum_j \frac{(\boldsymbol{n}_j \cdot (\boldsymbol{r}_j - \boldsymbol{R}_i))S_j}{|\boldsymbol{r}_j - \boldsymbol{R}_i|^n} \right]^{1/n-3} \quad n \geq 4 .$$

(36)

Мы ограничивались четырьмя интегралами $n$=4,5,6,7. Эмпирически подобраны значения параметров (для размерности расстояния в А):

$A_0$= -8.191 1/A, $A_1$=152.661, $A_2$= -238.555, $A_3$= 151.700, $A_4$= -1.074

### 2.1.3.2 Аналитические градиенты энергии SGB

Полный дифференциал энергии при неизменных зарядах и диэлектрических постоянных:

$$\frac{\partial \Delta G_{pol}}{\partial R_{m,x^k}} = -\frac{1}{2}\cdot\frac{1}{\varepsilon_{in}}\frac{1}{1+\frac{1}{2\varepsilon}}\left(1-\frac{1}{\varepsilon}\right)\cdot\sum_{i,j} Q_i\cdot Q_j\cdot\frac{\partial g_{i,j}}{\partial R_{m,x^k}},$$

(37)

где функция $g_{i,j}$ и ее производные определяются формулами:

$$g_{i,j} = \frac{1}{\sqrt{\left|\vec{R}_{i,j}\right|^2 + a_i\cdot a_j\cdot exp(-\frac{\left|\boldsymbol{R}_{i,j}\right|^2}{c\cdot a_i\cdot a_j})}},$$

(38)

$$\frac{\partial g_{i,j}}{\partial R_{m,x^k}} = \frac{-g_{i,j}^{\;3}}{2}\left[\begin{array}{l} 2((R_{i,x^k} - R_{j,x^k})\cdot(\delta_{im}-\delta_{jm})\left(1-\frac{1}{c}exp(-\frac{\left|\boldsymbol{R}_{i,j}\right|^2}{c\cdot a_i\cdot a_j})\right)+ \\ \left(\frac{\partial a_i}{\partial R_{m,x^k}}a_j + a_i\frac{\partial a_j}{\partial R_{m,x^k}}\right)\left(1+\frac{\left|\boldsymbol{R}_{i,j}\right|^2}{c\cdot a_i\cdot a_j}\right)exp(-\frac{\left|\boldsymbol{R}_{i,j}\right|^2}{c\cdot a_i\cdot a_j}) \end{array}\right].$$

(39)

Дифференциалы для борновских радиусов атомов:

$$\frac{\partial a_i}{\partial R_{m,x^k}} = \frac{-1}{2\left(\sum_n A_n\cdot I_n^i - A_0\right)^2}\cdot\sum_n A_n\cdot\frac{\partial I_n^i}{\partial R_{m,x^k}}.$$

(40)

Производные $I_n^i$ -интегралов (для дискретного приближения значений интегралов, $s$ – индекс поверхностного элемента):

$$\frac{\partial I_n^i}{\partial R_{m,x^k}} = \frac{\sum_s \frac{\partial J_{s,n}^i}{\partial R_{m,x^k}}}{n-3}\frac{1}{\left[I_n^i\right]^{n-4}} \quad n \geq 4,$$

(41)

где $J_{s,n}^{\;i}$ и их производные ищутся по формулам:

$$J^i_{s,n} = \frac{(\boldsymbol{n}_s \cdot (\boldsymbol{r}_s - \boldsymbol{R}_i))S_s}{\left|\boldsymbol{r}_s - \boldsymbol{R}_i\right|^n},$$

(42)

$$\frac{\partial J^i_{s,n}}{\partial R_{m,x^k}} = \left[ \frac{\left[\left(\boldsymbol{n}_s \cdot (\frac{\partial \boldsymbol{r}_s}{\partial R_{m,x^k}} - \delta_{im}\boldsymbol{e}^k)\right) + \left(\frac{\partial \boldsymbol{n}_s}{\partial R_{m,x^k}} \cdot (\boldsymbol{r}_s - \boldsymbol{R}_i)\right)\right] S_s + \left(\boldsymbol{n}_s \cdot (\boldsymbol{r}_s - \boldsymbol{R}_i)\right)\frac{\partial S_s}{\partial R_{m,x^k}}}{\left|\boldsymbol{r}_s - \boldsymbol{R}_i\right|^n} - \frac{n\left(\boldsymbol{n}_s \cdot (\boldsymbol{r}_s - \boldsymbol{R}_i)\right)\left((\frac{\partial \boldsymbol{r}_s}{\partial R_{m,x^k}} - \delta_{im}\boldsymbol{e}^k) \cdot (\boldsymbol{r}_s - \boldsymbol{R}_i)\right)S_s}{\left|\boldsymbol{r}_s - \boldsymbol{R}_i\right|^{n+2}} \right].$$

(43)

## 2.2 Неполярная компонента энергии сольватации.

### 2.2.1 Основные уравнения для неполярная компоненты энергии

Для расчета неполярной части энергии берутся поверхностные элементы на SAS.

Для подсчета неполярной части энергии используем формулу Bordner, Cavasotto, Abagyan [20] :

$$E_{SAS}=\sum_j \sigma_j S_j + b,$$

(44)

где $j=1,\dots,N_s$ по всем поверхностным элементам на SAS с ненулевой площадью.

Для воды:

$\sigma_j = \sigma = 0.00378$ - в ккал/(моль А$^2$)

$b$=0.698 - в ккал/моль (45)

### 2.2.2 Градиенты неполярной составляющей энергии сольватации

Градиент энергии неполярной составляющей:

$$\frac{\partial \Delta G_{nonpol}}{\partial R_{m,x^k}} = \sum_{j=1}^{N} \sigma_j \frac{\partial s_j}{\partial R_{m,x^k}} .$$

(46)

Индекс $j=1,\ldots,N_g$ пробегает по всем поверхностным элементам на SAS с ненулевым градиентом.

## 3. Компьютерная проверка расчетов.

Была проведена компьютерная проверка найденных выше формул с помощью программы DISOLV [1-3, 7-8]:

### 3.1 Сравнение численных и аналитических градиентов для методов PCM, COSMO и SGB. (Рис.2)

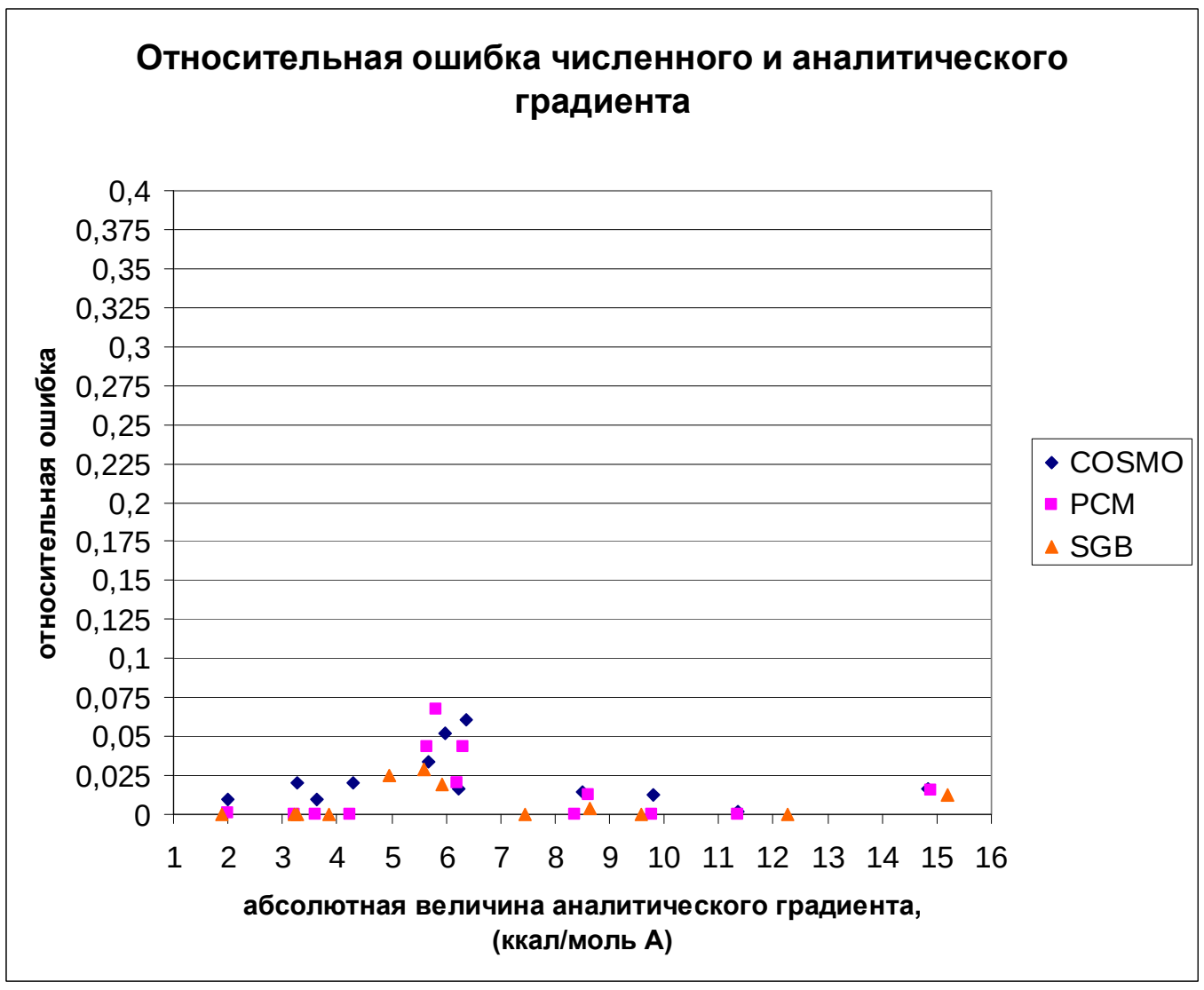

Рис. 2 Относительная ошибка численного и аналитического градиента для методов PCM, COSMO и SGB (для шага сетки 0.1) По оси x – величина численного  градиента энергии (COSMO или SGB), по оси y – относительная погрешность энергии по отношению к аналитическому градиенту (COSMO или SGB, соответствено): $\delta^{COSMO} = \frac{|\nabla_{\boldsymbol{R}_j} E_{num} - \nabla_{\boldsymbol{R}_j} E_{anal}|}{|\nabla_{\boldsymbol{R}_j} E_{num}|}$ .

Для метода PCM, COSMO и SGB получены следующие результаты по сравнению численных градиентов с аналитическими:

1) Для PCM:

относительная погрешность численной производной по отношению к аналитической для градиентов больших 2 ккал/(моль А) меньше 6,6 % и в среднем равна 2% Для объемных атомов погрешность численной производной по отношению к аналитической для градиентов больших 2 ккал/(моль А) меньше 0.03%.

2) Для SGB:

относительная погрешность численной производной по отношению к аналитической для градиентов больших 2 ккал/(моль А) меньше 3% и в среднем равна 1,5%. Для объемных атомов погрешность численной производной по отношению к аналитической для градиентов больших 2 ккал/(моль А) меньше 0.03%.

3) Для COSMO:

относительная погрешность численной производной по отношению к аналитической для градиентов больших 2 ккал/(моль А) меньше 6% и в среднем равна 3%

Для объемных атомов погрешность численной производной по отношению к аналитической для градиентов больших 2 ккал/(моль А) меньше 0.2%.

Ошибки численных градиентов объемных атомов определяются погрешностью вычисления энергии, а поверхностных атомов – прежде всего сдвигом сетки, происходящем при сдвиге атома.

### 3.2 Сравнение аналитических градиентов для COSMO и SGB с PCM

4. 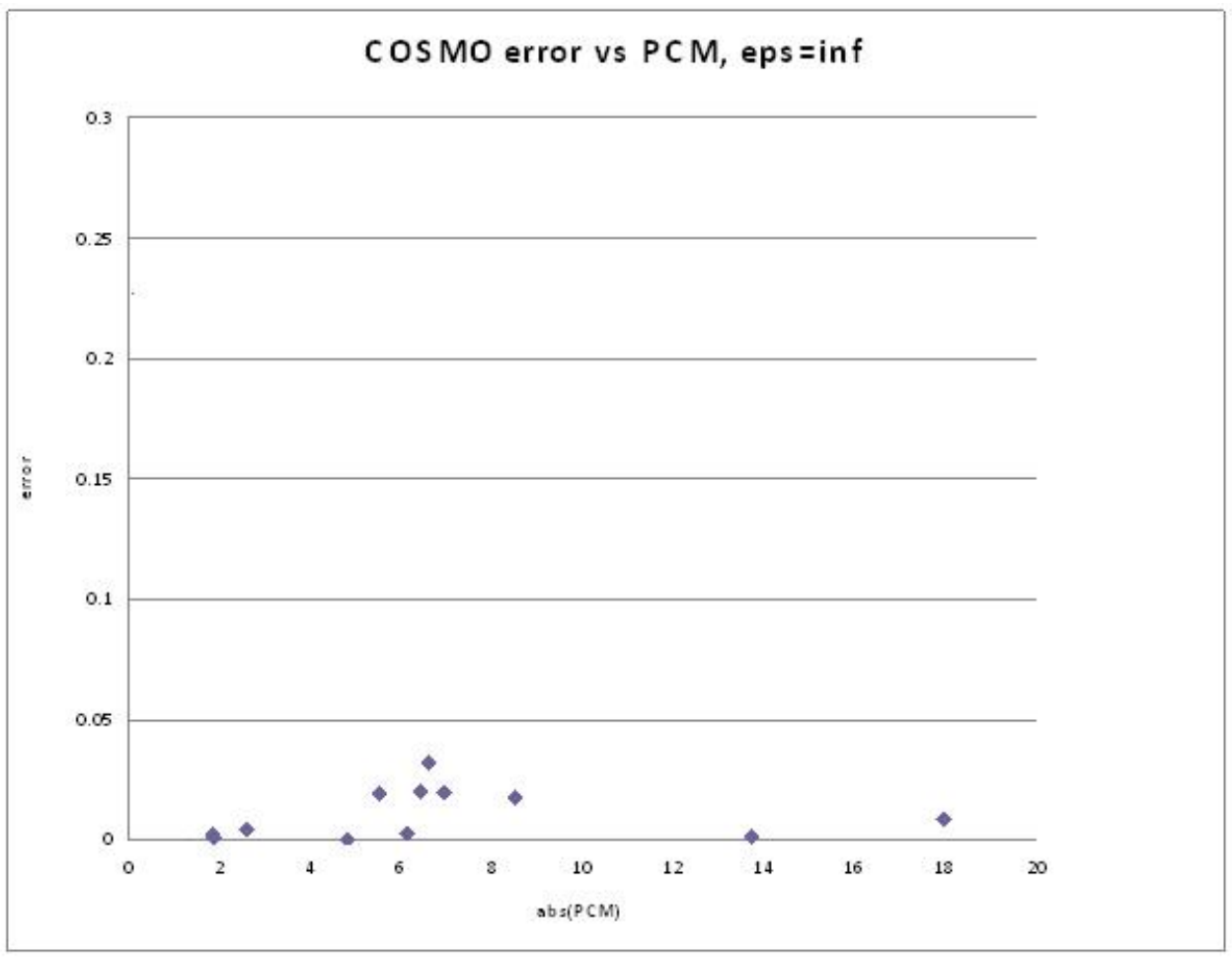

Рис. 3 Относительная ошибка аналитических градиентов COSMO в сравнении с PCM для $\varepsilon\to\infty$. По оси $x$ – величина градиента PCM, по оси $y$ – относительная погрешность по отношению к методу COSMO:

$$\delta = \frac{|\nabla_{\boldsymbol{R}_j} E_{COSMO} - \nabla_{\boldsymbol{R}_j} E_{PCM}|}{|\nabla_{\boldsymbol{R}_j} E_{PCM}|}.$$

1) аналитические градиенты для COSMO посравнению с PCM (Рис.3):

a) Для  диэлектрической проницаемости стремящейся к бесконечности $\varepsilon\to\infty$ и размера шага сетки стремящегося к нулю, аналитические градиенты для COSMO и PCM должны точно совпадать (т.к. COSMO является пределом PCM при бесконечной диэлектрической проницаемотси и бесконечно малом шаге сетки). Действительно, для COSMO для ε→∞ относительная погрешность аналитической производной по отношению к аналитической производной для PCM и для градиентов больших 2 ккал/(моль A) меньше 2.5% и в среднем равна 1.2%.  Максимальные значения относительной погрешности производной соответствует поверхностным атомам.

b) Для COSMO для ε=78.5 (вода) относительная погрешность аналитической производной по отношению к аналитической производной для PCM немного выше, а именно, для градиентов больших 2 ккал/(моль A) меньше 4% и в среднем равна 2%.

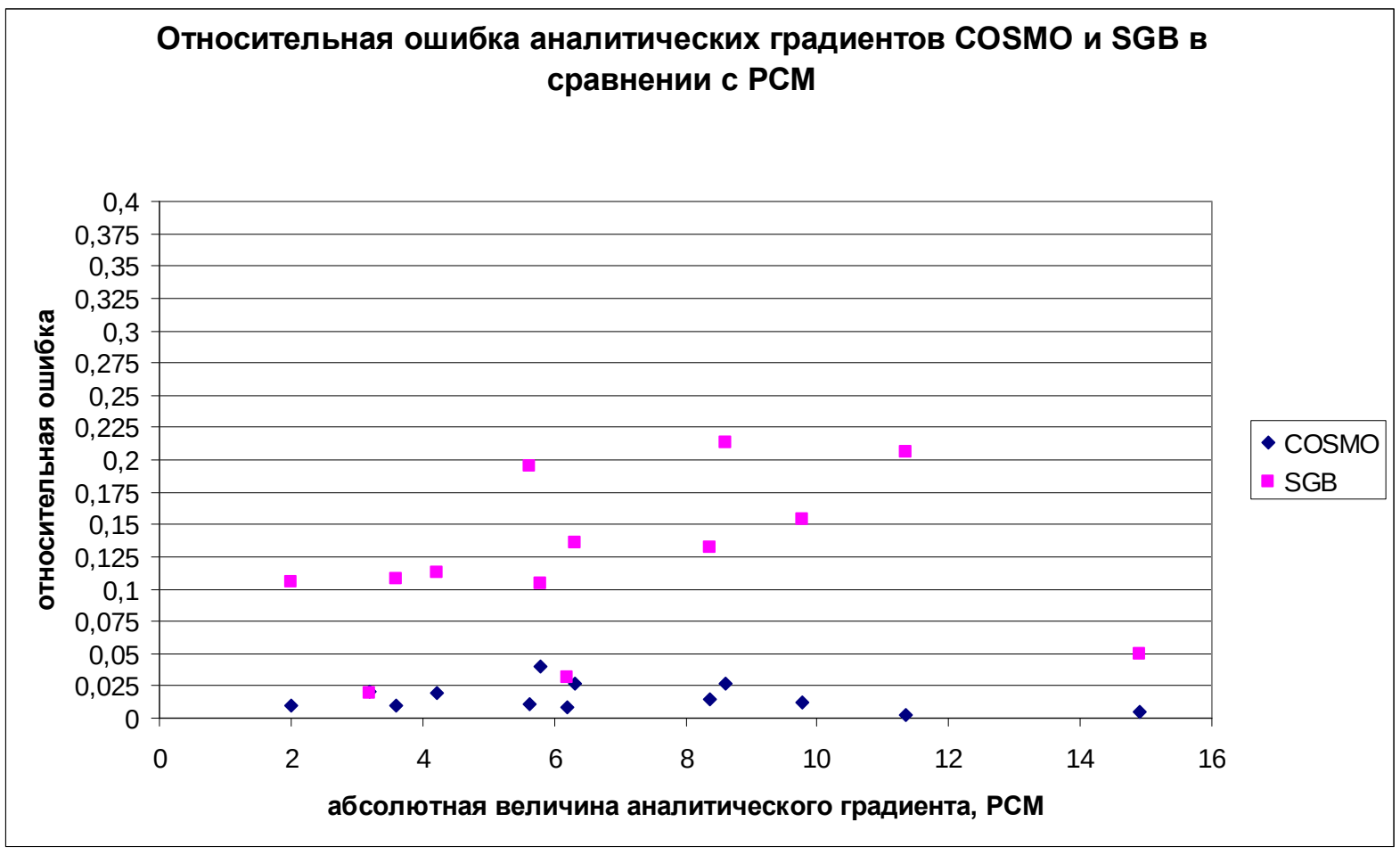

Рис. 4 Относительная ошибка аналитических градиентов COSMO и SGB в сравнении с PCM для *ε*=78.5. По оси *x* – величина аналитического градиента PCM, по оси *y* – относительная погрешность по отношению к методу (COSMO или SGB): $\delta = \dfrac{|\nabla_{R_j} E_{(COSMO,SGB)} - \nabla_{R_j} E_{PCM}|}{|\nabla_{R_j} E_{PCM}|}$.

2) Сравним аналитические градиенты для SGB с PCM (Рис.4):

Для SGB для ε=78.5 (вода) Относительная погрешность аналитической производной по отношению к аналитической производной для PCM много выше, а именно, для градиентов больших 2 ккал/(моль А) она меньше 23% и в среднем равна 13%. Это неудивительно, поскольку SGB это приближенная эмпирическая модель.

## Выводы.

В работе проведены расчеты аналитических градиентов энергии на основе методов PCM, COSMO и SGB для молекулярной поверхности, полученной методами первичной и вторичной обкатки в соответствии с алгоритмами, изложенными в [1-9].

Приведем проблемные случаи, которые могут возникать при расчете градиентов и решении задачи оптимизации геометрии молекул с их использованием:

- Вырождение - тройная точка опирается не на три, а на большее число атомов (например, ароматические кольца)
- Происходит резкая перестройка поверхности из-за изменения радиуса вторичной обкатки или критического расстояния самой программой при автоматической его настройке
- Происходит резкая перестройка поверхности из-за появления или исчезновении тора обкатки (первичного или вторичного) при сдвиге тройных точек или изменения узкого перешейка тора на расстояниях близких к критическому
- Происходит резкая перестройка поверхности из-за появления или исчезновении тройных точек первичных или вторичных.

Как результат могут быть скачки градиентов и «колебания» оптимизатора вблизи вышеописанных точек.

Были проведены компьютерные расчеты для подтверждения правильности формул для градиентов энергии, полученные разными методами. Хорошее совпадение аналитических градиентов между собой для разных методов, а также хорошее совпадение численных градиентов с аналитическими для метода COSMO подтверждает правильность найденных формул.

## Благодарности.

Данная работа является более детальным изложением и дальнейшим развитием статьи [9]. Авторы выражают глубокую признательность всем авторам этой публикации за проделанную работу, которая послужила основой и для данной статьи.

Список литературы